\documentclass[aps,prl,reprint,groupedaddress,showpacs]{revtex4-1}

\usepackage{amsmath}

\usepackage{amssymb}
\usepackage{amsthm}

\usepackage{braket}

\usepackage{graphicx}

\usepackage{color}

\begin{document}


\title{A general theory for irreversible thermodynamics}


\author{J. Ricardo Arias-Gonzalez}
\email[Corresponding author: ricardo.arias@imdea.org, \newline ricardo.ariasgonzalez@gmail.com]{}
\affiliation{Instituto Madrile\~{n}o de Estudios Avanzados en Nanociencia,
C/Faraday 9, Cantoblanco, 28049 Madrid, Spain}
\affiliation{CNB-CSIC-IMDEA Nanociencia Associated Unit
``Unidad de Nanobiotecnolog\'{i}a"}


\date{\today}

\begin{abstract}
We demonstrate that irreversibility arises from the principle of
microscopic reversibility and the presence of memory in the time evolution
of a single copy of a system driven by a protocol.
We introduce microscopic reversibility by using the concept of protocol- and
pathway-dependent thermodynamic function, as defined
in~\textit{J.R. Arias-Gonzalez, arXiv:1511.08017 [cond-mat.stat-mech]},
and memory by using the concept of non-Markovianity, as
in~\textit{J.R. Arias-Gonzalez, arXiv:1511.06139 [cond-mat.stat-mech]}.
We define work as the change in free energy and heat as the change in entropy
for micoscopic, individual pathways of a system subject to a protocol.
We find that all non-equilibrium statistics emerge naturally. In particular,
we derive most known fluctuation theorems and formulate two others.
While the conservation of energy is invoked both at the level of the 
individual pathway and in ensemble-average processes,
the second law of thermodynamics and the time arrow, which are only fulfilled
in ensemble-average processes, are shown to be
consequences of microscopic reversibility and non-Markovianity.\\
\end{abstract}


\maketitle


Non-equilibrium thermodynamics seeks to understand irreversible processes,
namely, those for which the initial state of a system becomes irrecoverable
without energy expenditure.
Statistical Mechanics, in turn, explains from first physical laws
how miscroscopic fluctuations lead to equilibrium processes in the macroscopic
limit~\cite{Chandler1987,Pathria2011}.
It is known that irreversible processes devaluate the energy of a
system, that is, part of the energy that a system exchanges with the
environment is wasted in the sense that it cannot be transformed into useful
work. This is the
second law of thermodynamics, which is obeyed by ensemble-averaged
(time-averaged if ergodicity can be applied) systems, not under an individual
pathway along which the system fluctuates sometimes generating so-called
transient violations of this law.

Fluctuation theorems have appeared in
recent years explaining quantitatively energy imbalances between
forward and reverse processes or between equilibrium and non-equilibrium
processes that connect an initially equilibrium state with a final, generally
non-equilibrium state of a system~\cite{Bustamante2005,Ritort2008}.
These theorems have been tested
experimentally~\cite{Liphardt2002,Collin2005,Douarche2005},
mostly in biomolecular systems analyzed on a one by one
basis~\cite{Bustamante2008}.
However, a fundamental theory based on general principles underlying those
theorems has not been established. In other words, there is to date no formal
framework that uses statistics and first physical laws to precisely
describe how irreversibility arises from fluctuations.

Thermodynamics establishes relationships among
ensemble-average values of the thermodynamic potentials, what ultimately
neglects the individuality of a single copy of the system
or the single pathway that the system describes.
Single-molecule biophysics is bringing to the foreground of the statistical
discussion the importance of single pathways and single systems because
biochemical reactions do not ocurr in ensembles in the
cell~\cite{Bustamante2008}.
It is then legitimized analyzing energy exchanges from the single object 
viewpoint.

We previously developed a general framework for microscopic reversible
pathways based on Non-Markovian relations between the
present and the previous states of a system and the fact
that it is fair to define thermodynamic functions for a
single copy of the system~\cite{Arias-Gonzalez2014a, Arias-Gonzalez2014b}.
Here, we use these concepts, namely,
(1), Microscopic reversibility, (2), Non-Markovianity and (3), what we may 
call here {\it individuality}, to explain non-equilibrium processes.
The third one is clear at least for classical systems, which are
distinguishable, and, as introduced, can be followed experimentally at the
single copy level. We will illustrate our theory by
deducing the most known fluctuation theorems and by introducing two new
closing relations.

We suppose that our system is in contact with a thermal bath at temperature $T$
at all times.
We recognize the existence of a protocol that stochastically drives the
evolution of the system through privileged pathways, each comprised of a set
of successive possible states according to a directional, stochastic
chain with memory~\cite{Arias-Gonzalez2014a}.
In the following, we will use the terms {\it substate} (or {\it event}),
{\it quasistate} and {\it state} as defined in our previous
work~\cite{Arias-Gonzalez2014b}:
substates will refer to the sequential stages that a system traverses in its
evolution, a quasistate will refer to a certain substate at time $t$ plus the 
history of substates that the system recalls and a state will refer to the
ensemble of quasistates
that comply with the constraints fixed by the protocol at time $t$.
In addition, here, we will use the term {\it pathway}
to refer to a single microscopic trajectory in the phase space,
as defined by a sequence of substates,
and will reserve the term {\it process} for the ensemble average over
the pathways that the system can follow under the existing protocol.
This ensemble average is different from a time average if the
memory of the system is long enough (i.e., if it comprises a sufficiently large
number of previous substates) with respect to the number of substates
that the system goes through in its evolution along a pathway under the
protocol.
In these conditions, the average over the ensemble of substates adopted by the
system along a certain pathway under a certain protocol may not be the same
that the average over an ensemble of copies of the system, each describing a
certain pathway driven by the same protocol, making the ergodic hypothesis not
valid in general.

We denote by $\lambda$ and $\lambda^{-1}$ the direct and inverse protocols,
respectively, and their associated set of control parameters.
The inverse protocol denotes the time-reversed forward protocol, hence it is
characterized by the same set of parameters with opposite evolution;
the superindex label ``$-1$'' may be then dropped down when there is no
confusion on time direction.
Given two temporal instants $t>\tau$,
we associate a directional, stochastic chain with
memory~\cite{Arias-Gonzalez2014a}, $\nu$, to a pathway, as defined by a
temporal sequence of substates, that shares a
common segment and that has been constructed under protocol $\lambda$ as:
\begin{eqnarray}\label{eq:nutlambda}
\nu_{t}^{(\lambda)} & = &
     \{x_0^{(\lambda)},\ldots,x_{\tau}^{(\lambda)},\ldots,x_t^{(\lambda)}\},
\\ [+1mm]
\label{eq:nutaulambda}
\nu_{\tau}^{(\lambda)} & = &
     \{x_0^{(\lambda)},\ldots,x_{\tau}^{(\lambda)}\}.
\end{eqnarray}
\noindent
When there is no confusion about the protocol, we will drop the
superindex $\lambda$.
Likewise, when there is no confusion by the time instants, we will drop
subindices $t$ and $\tau$ from the sequence index.

A thermodynamic function, ``$A$", can be averaged over the phase-space pathways 
that the system can follow complying with the evolving constraints
prescribed by the protocol $\lambda$ as~\cite{Arias-Gonzalez2014b}
\begin{equation}\label{eq:lpotential}
A^{(\lambda)} \equiv \left \langle A^{(\lambda)}_{\nu} \right \rangle_{\lambda}
= \sum_{\nu=1}^N p_{\nu}^{(\lambda)} A_{\nu}^{(\lambda)},
\end{equation}
\noindent
where $p_{\nu}^{(\lambda)}$ is the probability distribution according to
protocol $\lambda$. Probabilities are normalized by their corresponding
sequence-dependent partition
function~\cite{Arias-Gonzalez2014a,Arias-Gonzalez2014b}:
\begin{equation}\label{eq:lprob}
p_{\nu}^{(\lambda)} = \frac{e^{-\beta E_{\nu}}}{Z_{\nu}^{(\lambda)}},
\end{equation}
\noindent
such that $\sum_{\nu=1}^N p_{\nu}^{(\lambda)} =1$ and $\beta = 1/ k T$.
As explained ealier~\cite{Arias-Gonzalez2014a,Arias-Gonzalez2014b},
the standard partion function, $Z$, which
defines the equilibrium probabilities
$p_{\nu}= \exp \left( -\beta E_{\nu} \right) / Z$, does not make any
assumptions on a particular protocol, it comprises all the possibilities
for all the protocols and therefore it is independent of time.
Time, according to this scheme, can be conceived as implicitly determined,
on the one hand, by the number of states, $n$, that the system adopts between
$\tau$ and $t$ ($n$ is the cardinality of $\nu_t$ minus that of $\nu_{\tau}$,
namely, $n=|\nu_t|-|\nu_{\tau}|=t-\tau$) and,
on the other hand, by the protocol $\lambda$.
We may then formally express $t=t(n, \lambda)$.

Ensemble-average equilibrium thermodynamic functions are describable
in a similar fashion within this formalism as
$A \equiv \left \langle A_{\nu} \right \rangle
= \sum_{\nu=1}^N p_{\nu} A_{\nu}$.
It is important to note that if ``$A$" is a thermodynamic potential,
$A_{\nu}^{(\lambda)}$ and $A^{(\lambda)}$ are not state functions
in the sense of ensemble-average processes because their
values are both pathway- and protocol-dependent for $A_{\nu}^{(\lambda)}$
and protocol-dependent for $A^{(\lambda)}$.
Within the framework of the microcanonical and canonical ensembles, we will
characterize the system by the Internal Energy, ``$U$", the Helmholtz Free
Energy, ``$F$", and the Entropy, ``$S$", thermodynamic potentials using their
protocol- and pathway-dependent definitions~\cite{Arias-Gonzalez2014b}.

We introduce the reversible, microscopic work between two states
$x_t$ and $x_{\tau}$ along sequence $\nu$ under protocol $\lambda$ as the
Helmholtz free energy difference:
\begin{equation}\label{eq:RevWork}
W^{(\lambda)}_{\nu} (t,\tau) \equiv
      F^{(\lambda)}_{\nu} (t) - F^{(\lambda)}_{\nu} (\tau).
\end{equation}
\noindent
We can associate to each sequence $\nu$ an inverse counterpart $\nu^{-1}$,
which mathematically reads
$\nu^{-1} = \{x_t,x_{t-1}\ldots,x_{\tau+1},x_{\tau},x_{\tau-1},
\ldots,x_1,x_0\}$. The associated reversible work is
\begin{equation}\label{eq:InvWork}
W^{(\lambda^{-1})}_{\nu} (\tau, t) =
      F^{(\lambda^{-1})}_{\nu} (\tau) - F^{(\lambda^{-1})}_{\nu} (t) =
-W^{(\lambda)}_{\nu} (t, \tau).
\end{equation}
\noindent
It is indeed possible to return to the initial substate $x_{\tau}$ from the
final substate $x_t$ along many pathways. Let $\nu'$ be a sequence that shares
a common path with $\nu$ until $\tau$ and differs from it between
$\tau$ and $t$, namely,
$\nu' = \{x_0,x_1,\ldots,x_{\tau-1},x_{\tau},x'_{\tau+1},\ldots,
x'_{t-1},x_t\}$. Then,
$W^{(\lambda^{-1})}_{\nu'} (\tau, t) =
      F^{(\lambda^{-1})}_{\nu'} (\tau) - F^{(\lambda^{-1})}_{\nu'} (t) \neq
-W^{(\lambda)}_{\nu} (t,\tau)$, the equality holding when there are no
interactions with previous events~\cite{Arias-Gonzalez2014a}.

The equilibrium, ensemble-average (macroscopic) work reads
$W (t,\tau) = F (t) - F(\tau)$.
It is also possible to define the protocol-dependent ensemble-average work as
$W^{(\lambda)} (t,\tau) = F^{(\lambda)} (t) - F^{(\lambda)} (\tau)$, which
follows from
$W^{(\lambda)} = \left \langle W_{\nu}^{(\lambda)} \right \rangle_{\lambda}$
by using Eq.~(\ref{eq:lpotential}).

We now set the heat from the definition of entropy in macroscopic,
reversible processes. Equilibrium thermodynamics relates the differential
entropy in a reversible process at constant temperature $T$
to the change in heat divided by the temperature, i.e.
$ dS \equiv \delta Q /T$, where $\delta$ stands for inexact differential.
For a macroscopic, reversible cycle, $\Delta S=0$ and the heat exchanged with
the environment equals zero.
For an irreversible cycle, the Clausius theorem expresses the second
law of Thermodynamics as $\oint \delta Q/T <0$~\cite{Landau1980}.

Following our previous work~\cite{Arias-Gonzalez2014b}, we will assume that
all evolutions of the system at the single pathway level are microscopically
reversible.
For reasons that will become clearer later, we will use the term
{\it non-equilibrium} (or {\it irreversible}) for a process that is
protocol-dependent, preserving the term {\it equilibrium} when the process
does not dependent on a protocol, as explained above.
Clausius theorem can be extended to a general irreversible
process that connects an initial state, $1$, and a final state, $2$,
by splitting an irreversible cycle $1 \rightarrow 2 \rightarrow 1$ 
into a forward non-equilibrium process $1 \rightarrow 2$
with entropy change
$\Delta S^{(\lambda)} \equiv S_2^{(\lambda)} - S_1^{(\lambda)}=
\int_1^2 \delta Q^{(\lambda)}/T$
and a backward equilibrium process $2 \rightarrow 1$ with
entropy change $-\Delta S \equiv S_1 - S_2 = \int_2^1 \delta Q / T$.
Then, it follows that $\Delta S >  \Delta S^{(\lambda)}$, in accord with the
extended version of Clausius theorem~\cite{Landau1980}.
For a system in contact with a thermal bath, it is clear that
$T \Delta S > Q^{(\lambda)}$, which expresses that part of the heat generated
by the system is wasted and released into the environment.
If the system is isolated, the internal energy is conserved, $\Delta U=0$,
which means that the system does not exchange heat or work with the
environment, only internal transformations are allowed.
Transformations are thus adiabatic, $Q^{(\lambda)}=0$, and then $\Delta S >0$.
This is the most known expression of the second law of Thermodynamics,
which states that the entropy of an isolated system either increases
(irreversible process) or is zero (reversible process).

By applying the principle of microscopic
reversibility to a time-directional, stochastic chain with memory, the heat
that the system exchanges with the environment at constant temperature, $T$, set
by the thermal bath reads:
\begin{equation}\label{eq:RevHeat}
Q^{(\lambda)}_{\nu} (t,\tau) \equiv
      T \left ( S^{(\lambda)}_{\nu} (t) - S^{(\lambda)}_{\nu} (\tau) \right ).
\end{equation}
\noindent
Likewise, the protocol-dependent ensemble-average
heat and the equilibrium heat are
$Q^{(\lambda)} (t,\tau) =
     T \left(S^{(\lambda)} (t) - S^{(\lambda)} (\tau) \right)$
and
$Q (t,\tau) = T \left( S(t) - S(\tau) \right)$, respectively.
The energy conservation imposes that
$\Delta E_{\nu}^{(\lambda)} (t,\tau) =
W_{\nu}^{(\lambda)} (t,\tau) + Q_{\nu}^{(\lambda)} (t,\tau)=
\Delta F_{\nu}^{(\lambda)} (t,\tau) + T \Delta S_{\nu}^{(\lambda)} (t,\tau)$
for single temporal trajectories.
These relations are formally the same
for protocol-dependent ensemble-averages and equilibrium quantities.
We consider positive both the heat absorbed by the system
and the work supplied to the system. 

We next derive Jarzynski's equality~\cite{Jarzynski1997} from our formalism.
We consider a system that is initially in an equilibrium state $x_{\tau}$,
namely, it is conformed by an ensemble of statistically similar states,
and that evolves to a final state $x_t$,
which might be non-equilibrium or achieve equilibrium afterwards,
through irreversible pathways controlled by a protocol $\lambda$.
The equilibrium Helmholtz free energy difference, $\Delta F$, between
state $x_{\tau}$ and the equilibrium equivalent of the final state was shown
to be related to the work, $W$, needed to drive the transition by
``$\left \langle \exp(-\beta W) \right \rangle = \exp(-\Delta F)$".

To better understand this picture, we will consider the example of 
DNA replication, in which the initial state at $\tau$ is an ensemble of free,
non-interacting deoxyribonucleotide triphosphates (dNTP), which represents a
{\it macroscopic} equilibrium similar to that of an ideal gas.
The final quasistate at $t$ is in turn a linear arrangement of
deoxyribonucleotide monophosphates (dNMP) directionally-constructed by a
DNA polymerase (DNAp) by sequentially arranging the dNMPs according to a
prescribed DNA template ({\it single-molecule} case~\cite{Bustamante2008}).
The final state corresponds to the ensemble of directionally-constructed
stochastic chains of dNMPs ({\it bulk} case~\cite{Bustamante2008}),
each chain sequentially assembled by a DNAp according to the
prescribed DNA template~\cite{Bustamante2011,Arias-Gonzalez2012}.
The final equilibrium state represents the spontaneous process in which
dNMPs are non-enzymatically arranged on the template, which is a similar
case to the spin chain construction studied by Ising and which can be
approximated by an enzymatically-driven polymerization with proofreading
at infinitely slowly replication rates~\cite{Arias-Gonzalez2012}.

The Helmholtz free energy of the initial state
is a fixed parameter that we will denote $F (\tau)$.
We want to calculate the free energy difference $\Delta F=F(t) - F(\tau)$,
where $F (t)$ is the equilibrium free energy of the final state,
i.e. $F(t)=-(1/ \beta)\ln Z(t)$,
by following microscopically reversible trajectories instead of obtaining it
from the equilibrium thermalization, which is either an inexistent process
(polymerization is in reality a protein-mediated process) or needs an
infinite amount of time despite its spontaneity.
The change in free energy for each chain constructed under a protocol will be
denoted by
$\Delta F_{\nu}^{(\lambda)} (t) \equiv F_{\nu}^{(\lambda)} (t) - F (\tau)$.

\begin{figure*}[!ht]
\begin{center}
\includegraphics[width=\textwidth]{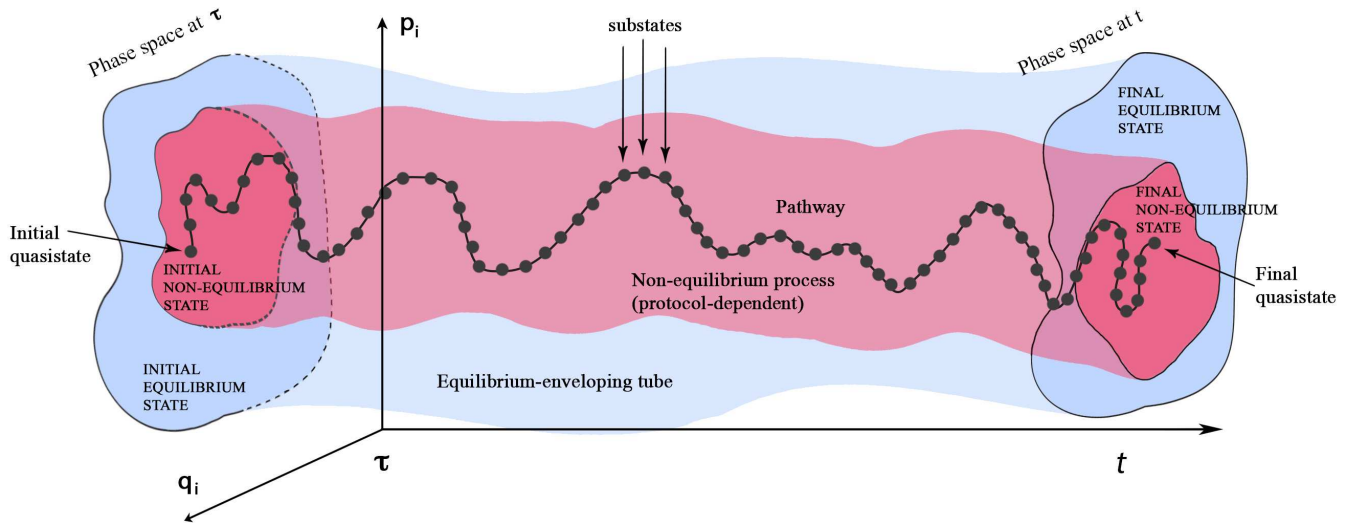}
\end{center}
\caption{{\bf Extended phase space.}
The scheme represents a reduced $(2D +1)$-dimensional
space of $2D$ generalized space and momentum coordinates,
$\left( q_h, p_h \right)$, $h=1,\ldots,D$ ($D$ is the number of degrees of
freedom), plus a proper time coordinate
(see the text for details). The system starts its evolution in an initial
quasistate at $\tau$ and ends in a final quasistate at $t$.
Blue hypersurfaces orthogonal to the temporal coordinate correspond to
equilibrium states and red ones to non-equilibrium states;
the extent and shape of these $2D$-dimensional hypersurfaces represent the
widths of the probability distributions at a definite height.
The red-shaded tube represents a non-equilibrium process,
which is the ensemble average over the individual pathways (black curve)
that the system can follow under protocol $\lambda$ by
traversing individual substates (black dots). The blue-shaded tube represents
the equilibrium process, which envelopes the equilibrium-probability
hypersurfaces between the two proper time instants.
Although a global time elapses along every pathway that the system describes,
the system does not age when pathways are orthogonal to the proper time
coordinate.
}
\label{fig:ExtPhaSpace}
\end{figure*}

Microscopically reversible transformations impose that
$W_{\nu}^{(\lambda)} = \Delta F_{\nu}^{(\lambda)}$. Then
\begin{eqnarray}\label{eq:expSmallWork}
\left \langle e^{-\beta \Delta F_{\nu}^{(\lambda)}}\right \rangle_{\lambda}
& = &
\left \langle e^{-\beta \left( F_{\nu}^{(\lambda)}-F(\tau) \right)}
\right \rangle_{\lambda}=
\left \langle \frac{Z_{\nu}^{(\lambda)}(t)}{Z(\tau)} \right \rangle_{\lambda} =
\frac{Z(t)}{Z (\tau)}
\nonumber \\ [+1mm]
& = &
e^{-\beta \Delta F},
\end{eqnarray}
\noindent
where we have used
$\left\langle Z_{\nu}^{(\lambda)} (t) \right\rangle_{\lambda} = Z(t)$,
as demonstrated elsewhere~\cite{Arias-Gonzalez2014a,Arias-Gonzalez2014b}.
It is important to note that the ensemble-average final free energy is
$F^{(\lambda)} \equiv \left \langle F_{\nu}^{(\lambda)}
\right \rangle_{\lambda} \neq F$.
What is actually true is the inequality~\cite{Arias-Gonzalez2014b}
\begin{equation}\label{eq:Fineq}
F^{(\lambda)} (t) \geq F(t),
\end{equation}
\noindent
which means that the reversible work to build an ensemble of directional,
stochastic chains with memory by microscopically reversible trajectories,
$W^{(\lambda)}=\Delta F^{(\lambda)}$, always supersedes that of the ensemble of
chains with memory that results from an equilibrium thermalization,
$W=\Delta F$:
\begin{equation}\label{eq:WorkIneq}
W^{(\lambda)} \geq W.
\end{equation}
\noindent
Inequality~(\ref{eq:WorkIneq}) can also be deduced from
Eq.~(\ref{eq:expSmallWork}), as formerly shown by
Jarzynski~\cite{Jarzynski1997}, by using the fact that the exponential is a
convex function, namely,
\begin{eqnarray}\label{eq:demoexpSmallWork}
\left \langle e^{-\beta \Delta F_{\nu}}\right \rangle_{\lambda} & \geq &
e^{-\beta \left \langle \Delta F_{\nu} \right \rangle_{\lambda}}=
e^{-\beta \Delta F^{(\lambda)}}
\nonumber \\ [+1mm]
& \Longrightarrow &
\Delta F^{(\lambda)} \geq \Delta F.
\end{eqnarray}
\noindent
Inequalities~(\ref{eq:WorkIneq}) and~(\ref{eq:demoexpSmallWork})
state that when memory is present,
the ensemble-average work employed to construct a stochastic chain of events
through microscopically reversible trajectories is always greater than the
process that involves an equilibrium thermalization.

Next, we derive Crooks theorem~\cite{Crooks1999}, which relates the
transition probabilities between microscopically reversible forward and
backward pathways. The system, as in Jarzynski equality, starts in an
equilibrium state. More in depth, it starts in one substate, $x_{\tau}$, out of
an ensemble of substates,
$\nu_{\tau} = \{x_{\tau(1)},\ldots,x_{\tau(m_0)}\}$,
that have been adopted under equilibrium conditions.
We use subindices $\tau(k)$, $k=1,\ldots,m_0$ to note that time is not
defined when the system experiences transitions in phase space in equilibrium.
Then, the system ends up in a general non-equilibrium substate,
$x_t^{(\lambda)}$, which might eventually
thermalize into a final equilibrium substate, $x_t$. As previously, $x_t$
represents one substate out of an ensemble of substates
$\nu_t = \{x_{t(1)},\ldots,x_{t(m_f)}\}$ (with $t(k)$, $k=1,\ldots,m_f$)
that have been achieved under protocol-independent pathways.

It is important to note that although time is not defined for a system that
experiencies equilibrium processes, a global time goes by from a macroscopic
point of view because other nearby systems involved in non-equilibrium
processes mark this global time arrow. More in depth, since equilibrium states
are idealizations because all systems are in the end interacting with one
another, a global time can be defined.
We will then use the term {\it proper time} to refer to the time coordinate of
a specific system, which, as explained, may pause if the system stays in
equilibrium with respect to the global time.

We introduce in Fig.~\ref{fig:ExtPhaSpace} the {\it extended phase-space},
which is the phase space representation that we believe suitable for both
equilibrium and non-equilibrium processes.
The scheme shows the propagation of the phase space of a system along its
proper time coordinate.
Hyperplanes orthogonal to the proper time coordinate represent
phase-spaces at each instant on which the system may explore equilibrium
configurations.
Hypersurfaces on individual phase-spaces correspond to states,
either equilibrium or non-equilibrium,
and the extent of the hypersurfaces on the hyperplanes represent the
functional profile of their probability distributions at a definite height,
either $p_{\nu}$ or protocol-dependent probabilities $p_{\nu}^{(\lambda)}$,
respectively.
The corrugated pipes represent processes, either equilibrium or
non-equilibrium, although rigorously speaking the former must exclusively sit
on hyperplanes perpendicular to the proper time coordinate.

Now, our system evolves from substate $x_{\tau}$ to $x_t^{(\lambda)}$ under the
influence of the protocol by following a pathway, namely,
$\nu_{\tau \rightarrow t}^{(\lambda)} = 
\{x_{\tau},x_{\tau+1}^{(\lambda)},\ldots,x_t^{(\lambda)}\}$,
see Eqs.~(\ref{eq:nutlambda}) and~(\ref{eq:nutaulambda}).
These conditions are applied for the reverse process, where now $x_t$ is the
equilibrium initial state and $x_{\tau}^{(\lambda^{-1})}$ is the general
non-equilibrium final state.
The transition forward probability, $P_F$, equals the
probability $p_{\nu} (\tau)$ of starting in equilibrium
times the probability, $p_{\nu}^{(\lambda)} (t)$, of ending in a
general state through a non-equilibrium pathway controlled by
protocol $\lambda$. The transition reverse probability, $P_R$, follows the
same rationale, namely,
$P_R = p_{\nu} (t) \times p_{\nu}^{(\lambda^{-1})} (\tau)$.
Then, the ratio of forward and reverse probabilities read
\begin{eqnarray}\label{eq:CrooksDemo1}
\frac{P_F}{P_R} & = & \frac{p_{\nu} (\tau) \times p_{\nu}^{(\lambda)} (t)}
                       {p_{\nu} (t) \times p_{\nu}^{(\lambda^{-1})} (\tau)}
\nonumber \\ [+1mm]
& = &
\frac{\exp \left(-\beta E_{\nu_{\tau}}\right)} {Z(\tau)}
\frac{\exp \left(-\beta E_{\nu_t}\right)} {Z_{\nu}^{(\lambda)}(t)}
\nonumber \\ [+1mm] 
& \times &
\left(
\frac{\exp \left(-\beta E_{\nu_t}\right)} {Z(t)}
\frac{\exp \left(-\beta E_{\nu_{\tau}}\right)} {Z_{\nu}^{(\lambda^{-1})}(\tau)}
\right)^{-1}
\nonumber \\ [+1mm]
& = &
\frac{Z(t)}{Z(\tau)}
\frac{Z_{\nu}^{(\lambda^{-1})}(\tau)}{Z_{\nu}^{(\lambda)}(t)}.
\end{eqnarray}
\noindent
The partition functions in Eq.~(\ref{eq:CrooksDemo1}) can be expressed in
terms of the free energies,
\begin{eqnarray}\label{eq:Feqttau}
Z(t) & = & \exp \left(-\beta F(t)\right),
\\ [+1mm]
Z_{\nu}^{(\lambda)} (t) & = & \exp \left(- \beta F_{\nu}^{(\lambda)}(t)\right);
\\ [+1mm]
Z(\tau) & = & \exp \left(-\beta F(\tau)\right),
\\ [+1mm]
Z_{\nu}^{(\lambda^{-1})} (\tau) & = &
         \exp \left(-\beta F_{\nu}^{(\lambda^{-1})}(\tau) \right).
\end{eqnarray}
\noindent
Then,
\begin{equation}\label{eq:CrooksDemo2}
\frac{P_F}{P_R} = \exp \left(
                  \frac{W_{\nu}^{(\lambda)} (t,\tau) - \Delta F}{k T}
                       \right),
\end{equation}
\noindent
where we have used the definition of the reversible microscopic work,
Eq.~(\ref{eq:RevWork}), and that $\Delta F = F(t) - F(\tau)$.

In the following, we will use our formalism to derive two new fluctuation
theorems for the relationship between the entropy and the heat exchanged
between the system and the environment in general, irreversible trajectories.
These theorems are the analogues of Crooks and Jarzynski theorems for
the relationship between the Helmholtz free energy and the work.
We express the protocol-dependent probability in terms of the entropy of a
single trajectory~\cite{Arias-Gonzalez2014b} as
\begin{equation}\label{eq:microCanonical}
p_{\nu}^{(\lambda)} = \exp \left ( 
                         - \frac{1}{k} S_{\nu}^{(\lambda)}
                           \right ).
\end{equation}

Like for the Crooks theorem, the system starts in an equilibrium state both
for the forward and reverse processes but this time the equilibrium is
defined under the microcanonical ensemble, that is, equilibrium states are
those with defined energy hence corresponding to the isolated
system. The non-equilibrium, forward pathways transform the system from an
equilibrium state with energy $U(\tau)$ into a non-equilibrium
quasistate with energy $E_{\nu} (t)$ belonging to a set that complies with an
internal energy $U^{(\lambda)} (t)$. This non-equilibrium state may eventually
thermalize into the corresponding equilibrium state with energy
$U (t) \geq U^{(\lambda)} (t)$, that is, the final energies $E_{\nu} (t)$
($\nu =1,\ldots,N$), 
may relax either by redistributing into equal degenerate energies $U (t)$
(so that $U (t) = U^{(\lambda)} (t)$) or the system may just become a subsystem
of a bigger system that contains the bath (so that $U (t) > U^{(\lambda)} (t)$).
In the reverse process, we similarly conceive that the system starts in this
equilibrium state and transforms into a non-equilibrium state with energy
$U^{(\lambda^{-1})} (\tau)$,
which may eventually relax into an equilibrium state with energy $U (\tau)$. 

The transition forward microcanonical probability, $p_F$, equals the
probability
$\exp \left(-S (\tau) / k \right)$ of starting in equilibrium
times the probability, $\exp \left (-S^{(\lambda)}_{\nu} (t) / k \right)$,
of ending in a general state through a non-equilibrium pathway controlled by
protocol $\lambda$. The transition reverse probability, $p_R$, follows the
same rationale, namely,
$p_R = \exp \left(-S (t)/k \right)
\times \exp \left(-S^{(\lambda^{-1})}_{\nu} (\tau)/k \right)$.
Then, the ratio of forward and reverse probabilities read
\begin{eqnarray}\label{eq:AriasFT1demo}
\frac{p_F}{p_R} & = &
\exp \left(  \frac{S (t) - S (\tau)} {k} \right)
\nonumber \\ [+1mm]
& \times &
\exp \left(
-\frac{S^{(\lambda)}_{\nu}  (t) - S^{(\lambda^{-1})}_{\nu} (\tau)}{k}
\right).
\end{eqnarray}
\noindent
Using the definition of the reversible microscopic heat,
Eq.~(\ref{eq:RevHeat}), the first, new fluctuation theorem states:
\begin{equation}\label{eq:AriasFT1}
\frac{p_F}{p_R} =
\exp \left( \frac{\Delta S}{k} -
            \frac{Q_{\nu}^{(\lambda)} (t,\tau)}{k T} \right).
\end{equation}
\noindent
Theorem~(\ref{eq:AriasFT1}) relates the heat exchanged between the system and
the environment under general, non-equilibrium processes to the ratio of the
forward and reverse probabilities. It is important to note that the system
is in contact with a thermal bath at temperature $T$ and therefore, the heat
$Q_{\nu}^{(\lambda)}$ is that generated by the system when it connects an
equilibrium state with defined energy to a generally non-equilibrium state.
Equation~(\ref{eq:AriasFT1}) can be simply expressed as
``$\frac{P_F (Q)}{P_R (-Q)} = \exp \left( \frac{\Delta S}{k} -
\frac{Q}{kT} \right)$",
as the counterpart expression of Crooks theorem appeared elsewhere
(see, for example~\cite{Bustamante2005}),
keeping in mind that $Q$ is the heat exchanged over irreversible paths,
in contrast to the notation used in this paper, where $Q$ is the
reversible heat, $Q \equiv T \Delta S$.

Like for the derivation of Jarzynski equality from Crooks
theorem~\cite{Crooks1999}, that is, by summing over the heat on both sides
of Eq.~(\ref{eq:AriasFT1}), it is easy
to obtain the second, new fluctuation theorem in this paper:
\begin{equation}\label{eq:AriasFT2}
\left \langle \exp \left( \frac{Q_{\nu}^{(\lambda)}}{k T}
              \right) \right \rangle_{\lambda} =
              \exp \left( \frac{\Delta S}{k} \right),
\end{equation}
\noindent
which may be invoked as
``$\left \langle \exp \left( \frac{Q}{k T} \right) \right \rangle =
              \exp \left( \frac{\Delta S}{k} \right)$"
keeping in mind that the expected value must be taken
in experiments always driven by the same protocol between an initially
equilibrium state with defined internal energy and a generally non-equilibrium
state with final energy $E_{\nu}$, and that $Q$ is a generally
irreversible heat.

Like for the demonstration of Jarzynski theorem, Eq.~(\ref{eq:expSmallWork}),
we can show how Eq.~(\ref{eq:AriasFT2}) naturally arises from our formalism.
Namely, microscopically reversible transformations impose that
$Q_{\nu}^{(\lambda)} = T \Delta S_{\nu}^{(\lambda)}$,
see Eq.~(\ref{eq:RevHeat}). Then:
\begin{eqnarray}\label{eq:expSmallHeat}
\left \langle e^{\frac{1}{k} \Delta S_{\nu}^{(\lambda)}}\right \rangle_{\lambda}
& = &
\left \langle e^{\frac{1}{k} \left( S_{\nu}^{(\lambda)} (t) - S (\tau) \right)}
\right \rangle_{\lambda}=
\left \langle \frac{p_{\nu} (\tau)}{p_{\nu}^{(\lambda)}} \right
\rangle_{\lambda} = \frac{p_{\nu} (\tau)}{1/N}
\nonumber \\ [+1mm]
& = &
e^{\frac{1}{k} \Delta S},
\end{eqnarray}

\noindent
where we have used that $N$ is the number of configurations compatible
with a final energy $U^{(\lambda)} (t)$ which thermalizes into $U (t)$.

Fluctuation theorem~(\ref{eq:AriasFT2}) relates the heat exchanged between the
system and the environment
between two states to their entropy difference $\Delta S$.
Using Jensen's inequality, it follows from Eq.~(\ref{eq:AriasFT2}) that
\begin{equation}\label{eq:Clausiusth}
\frac{1}{T} \left \langle Q_{\nu}^{(\lambda)} \right \rangle_{\lambda} =
\frac{Q^{(\lambda)}}{T} \leq \frac{Q}{T} = \Delta S,
\end{equation}
\noindent
which is the extended version of the Clausius theorem~\cite{Landau1980}
at constant temperature $T$,
as described above (remember that here $Q$ is the heat exchanged in a
reversible process and $Q^{(\lambda)}$ is the heat exchanged in a general
process under protocol $\lambda$).
Inequality~(\ref{eq:Clausiusth}) is the heat-entropy
counterpart to inequality~(\ref{eq:WorkIneq}).

Inequality~(\ref{eq:Clausiusth}) can be expressed as
$\Delta S^{(\lambda)} \leq \Delta S$, which can directly be deduced
from~\cite{Arias-Gonzalez2014b}
\begin{equation}\label{eq:Sineq}
S(t) - S^{(\lambda)}(t) \geq \frac{1}{T} \left( U(t) - U^{(\lambda)}(t) \right),
\end{equation}
\noindent
by using the fact that $U - U^{(\lambda)} \geq 0$ for the final, equilibrium
state of the isolated system (see above).
Inequality~(\ref{eq:Sineq}) expresses in the end that the entropy of the
isolated system always increases (see above).
On the contrary, when the system is not isolated it can be taken to states
of lower or higher entropy through non-equilibrium processes by appropriately
favouring pathways (i.e., by using certain protocols $\lambda$).
In particular, the system may increase its entropy with respect to the
equilibrium level at the cost of energy absorption ($S-S^{(\lambda)} < 0$,
$U-U^{(\lambda)} < 0$), as was shown for a DNA replication protocol in which
dNTPs where directionally assembled~\cite{Arias-Gonzalez2012},
or may decrease its entropy at the cost of energy dissipation
($S-S^{(\lambda)} > 0$, $U-U^{(\lambda)} > 0$), as was also shown for the same
system in a protocol in which dNMPs were removable~\cite{Andrieux2008b}.
Note that in~\cite{Arias-Gonzalez2012}, the DNAp was considered a passive
element that couples dNTP hydrolysis to dNMP incorporation and strand
hybridization, but this protein is actually an active nanomachine that uses
part of the dNTP energy to fuel its motor,
which activity includes translocation and accurate nucleotide incorporation.
DNAp action can thus be rightly included in the protocol $\lambda$ that drives
polymerization to the real, high fidelities~\cite{Bustamante2011,Morin2015}.

The lower the entropy of the final state (or the information acquisition) the
larger the energy dissipation~\cite{Andrieux2008b}, which, as explained above,
can be achieved in DNA replication by more complex protocols than the simply
passive, directional mechanism proposed in~\cite{Arias-Gonzalez2012}.
Like for the dissipated work,
$W_{diss}^{(\lambda)} \equiv W^{(\lambda)} - \Delta F$
(see, for example,~\cite{Jarzynski1997}), it is possible to define the
dissipated heat,
$Q_{diss}^{(\lambda)} \equiv - \left( Q^{(\lambda)} - T \Delta S \right)$,
which better illustrates this tradeoff.
$Q_{diss}^{(\lambda)}$ is the heat that cannot be used to decrease the entropy
of the system  and that is eventually realeased to the environment.

We now introduce the rate, $\sigma^{(\lambda)}$, at which the system exchanges
heat with the bath under a certain protocol $\lambda$:
\begin{equation}\label{eq:ExHeat}
\sigma^{(\lambda)}\equiv \frac{1}{T \Delta t } Q_{diss}^{(\lambda)}
= - \frac{Q^{(\lambda)} - Q}{T \Delta t} =
- \frac{\Delta S^{(\lambda)} - \Delta S}{\Delta t},
\end{equation}
\noindent
where $\Delta t \equiv t - \tau$ is the time interval.
With this definition, it is straightforward to derive the
Gallavotti-Cohen expression~\cite{Gallavotti1995}, which emerges from the
so-called {\it Fluctuation Theorem}~\cite{Evans1994,Evans1993}.
This theorem connects the forward and reverse
heat-rate probability distributions for steady-state, non-equilibrium
processes. It is demonstrated by following a similar argument as the one
we used for Eq.~(\ref{eq:AriasFT1}),
namely, by using Eq.~(\ref{eq:ExHeat}) for steady state conditions
in the derivation of Eq.~(\ref{eq:AriasFT1}), it follows that
\begin{equation}
\lim_{\Delta t \to \infty} \frac{k}{\Delta t} \ln \left( \frac{p_f}{p_r}\right)
= \sigma^{(\lambda)},
\end{equation}
\noindent
where
$p_f = \exp \left(-S (\tau)/k \right)
\times \exp \left(-S^{(\lambda)} (t)/k \right)$
and
$p_r = \exp \left(-S (t)/k \right)
\times \exp \left(-S^{(\lambda^{-1})} (\tau)/k \right)$.
Here, $\exp \left(-S^{(\lambda)} (t)/k \right)$
and $\exp \left(-S^{(\lambda^{-1})} (\tau)/k \right)$ are the probabilities of
finding the system with steady-state energies
$U^{(\lambda)}(t)$ and $U^{(\lambda^{-1})}(\tau)$.
In a cycle, $Q_{diss}^{(\lambda)}= -Q^{(\lambda)}$ and
$\sigma^{(\lambda)} = -Q^{(\lambda)}/T \Delta t$, since the
reversible heat is zero, and the internal energy change of the system is
$U^{(\lambda)}(t) - U^{(\lambda^{-1})}(\tau)$. The fact that
$\sigma^{(\lambda)} = -Q^{(\lambda)}/T \Delta t$ in a closed loop can also be
understood from the fact  that the system does not necessarily returns to the
initial state through the same pathway or with the inverse protocol.
\\

We have used microscopic reversibility, non-Markovianity and the fact that
a system can be experimentally followed at the single copy level to explain
that irreversibility arises from the average over the ensemble of pathways that
the system can follow in its evolution under time-dependent constraints.
Just by using conservation laws it is possible to consider that no temporal
evolution of a system is irreversible at the single-pathway level
and to deduce irreversibility as a consequence of protocol-biased
stochasticity in the presence of memory.
A system may be then said to evolve through a non-equilibrium
process when there are memory effects between each present substate and its
corresponding past substates along every available pathway driven by a
protocol. We have proposed extended phase-space diagrams to represent both
equilibrium and non-equilibrium processes.

In the absence of memory, all processes take place in equilibrium,
as was early demonstrated (see the Independence limit
theorem~\cite{Arias-Gonzalez2014a}); non-equilibrium is just a consequence
of the impossibility to explore all substates to connect arbitrary states in a
definite time.
In this regard, protocols that drive the system between two states at 
high velocities decrease the chances of the system to explore a significant
number of substates in each possible pathway, thus decreasing the probability
that the reverse pathways that may be used to recover the system become
similar to the time-reversed of the forward ones,
thus driving the system out of equilibrium.
Specifically, let's consider two identical protocols, $\lambda$ and
$\lambda'$, except for the velocity, being the latter faster than the former.
Then, protocol $\lambda$ allows the system to explore more substates within the
same microscopically reversible pathway to connect arbitrary quasistates than
$\lambda'$.
In a cycle, protocol $\lambda$ is more likely to drive the system in the
forward direction through pathways that can be aproximately mapped reversely
in the backward direction (by inverse protocol $\lambda^{-1}$)
than protocol $\lambda'$.
The system would then evolve farther from equilibrium by protocol $\lambda'$
than by protocol $\lambda$.
In the limit in which a protocol is infinitely slowly, the system visits
all substates to connect arbitrary states both for the forward and the
backward directions, hence making the process in equilibrium.

We have deduced Jarzynski, Crooks and Gallavotti-Cohen theorems and
have presented two closing fluctuation theorems that relate the heat
exchanged by a system when it transforms irreversibly between an equilibrium
state and a final, general quasistate.
We have shown that dissipated heat (or friction) appears as a consequence of
the unlikelyhood for the system to describe overlapping forward and backward
pathways.

The second law of Thermodynamics can be definitely observed as a consequence
of both microscopic reversibility and memory effects, as defined by stochastic
interactions among the substates that the system adopts along its
evolution. Therefore, it may not be considered as a fundamental law of Physics
but rather a consequence of conservation laws.
Note that unlike the first law of Thermodynamics,
the second law only arises after ensemble averages, whereas the energy
conservation is fulfilled at both individual pathways and in
ensemble-average processes. The time arrow may also be considered
a consequence of microscopic reversibility and non-Markovianity.
In fact, in equilibrium and in systems where memory effects can be
neglected, the time coordinate is not defined, which indicates that a system
does not age in these conditions. This makes possible to define a
proper time for each system, which indicates how the system ages.
The fact that systems are not isolated make proper times correlate into a
global time with which aging in different systems can be compared.
The proper time that elapses while a system evolves irreversibly between two
states is determined by the number of intermediate states that are gone
through under a defined protocol.

\noindent
\textbf{Acknowledgments}\\
P. S\'anchez Molina is thanked for illustration assistance in Fig. 1.
Work supported by IMDEA Nanociencia.


\begin{thebibliography}{21}
\bibitem{Chandler1987} Chandler, D. {\it Introduction to Modern Statistical Mechanics} (Oxford University Press, 1987).

\bibitem{Pathria2011} Pathria, R.~K. \& Beale, P.~D. {\it Statistical Mechanics (Third Edition)} (Academic Press, Boston 2011).

\bibitem{Bustamante2005} Bustamante, C., Liphardt, J. \& Ritort, F. The nonequilibrium thermodynamics of small systems. \textit{Physics Today}, \textbf{58}, 43-48 (2005).

\bibitem{Ritort2008} Ritort, F. Nonequilibrium fluctuations in small systems: From physics to biology. In: Rice, S.~A., editor. \textit{Adv. Chem. Phys.}, Wiley. Chapter 2, \textbf{137}, 31-123 (2008).

\bibitem{Liphardt2002} Liphardt, J., Dumont, S., Smith, S.~B., Tinoco, I., \& Bustamante, C. Equilibrium Information from Nonequilibrium Measurements in an Experimental Test of Jarzynski’s Equality. \textit{Science}, \textbf{296}, 1832-1835 (2002).

\bibitem{Collin2005} Collin, D., Ritort, F., Jarzynski, C., Smith, S.~B., Tinoco, I., \& Bustamante, C. Verification of the Crooks fluctuation theorem and recovery of RNA folding free energies. \textit{Nature}, \textbf{437}, 231-234 (2005).

\bibitem{Douarche2005} Douarche, F., Ciliberto, S., Petrosyan, A., \& Rabbiosi, I. An experimental test of the Jarzynski equality in a mechanical experiment. \textit{Europhys. Lett.}, \textbf{70}, 593-599 (2005).

\bibitem{Bustamante2008} Bustamante, C. In singulo Biochemistry: When Less Is More. \textit{Annu. Rev. Biochem.}, \textbf{77}, 45-50 (2008).

\bibitem{Arias-Gonzalez2014a} Arias-Gonzalez, J.~R. Statistical physics of directional, stochastic chains with memory. arXiv:1511.06139 [cond-mat.stat-mech]

\bibitem{Arias-Gonzalez2014b} Arias-Gonzalez, J.~R. A general framework for microscopically reversible processes with memory. arXiv:1511.08017 [cond-mat.stat-mech]

\bibitem{Landau1980} Landau, L.~D., \& Lifshitz, E.~M. {\it Statistical Physics}, Part 1, Sect. 13 (Pergamon Press, 1980).

\bibitem{Jarzynski1997} Jarzynski, C. Nonequilibrium Equality for Free Energy Differences. \textit{Phys. Rev. Lett.}, \textbf{78}, 2690-2693 (1997).

\bibitem{Bustamante2011} Bustamante, C., Cheng, C. \& Mejia, Y.~X. Revisiting the Central Dogma One Molecule at a Time. \textit{Cell}, \textbf{144}, 480-497 (2011).

\bibitem{Arias-Gonzalez2012} Arias-Gonzalez, J. R. Entropy involved in fidelity of DNA replication. \textit{PLoS ONE}, \textbf{7}, e42272 (2012).

\bibitem{Crooks1999} Crooks, G.~E. Entropy production fluctuation theorem and the nonequilibrium work relation for free energy differences. \textit{Phys. Rev. E}, \textbf{60}, 2721-2726 (1999).

\bibitem{Andrieux2008b} Andrieux, D. \& Gaspard, P. Nonequilibrium generation of information in copolymerization processes. \textit{Proc. Natl. Acad. Sci. USA}, \textbf{105}, 9516-9521 (2008).

\bibitem{Morin2015} Morin, J.~A., Cao, F.~J., L\'azaro, J.~M., Arias-Gonzalez, J.~R., Valpuesta, J.~M., Carrascosa, J.~L., Salas, M., \& Ibarra, B. Mechano-chemical kinetics of {DNA} replication: identification of the translocation step of a replicative {DNA} polymerase. \textit{Nucleic Acids Res.}, \textbf{43}, 3643-3652 (2015).

\bibitem{Gallavotti1995} Gallavotti, G., \& Cohen, E.~G.~D. Dynamical Ensembles in Nonequilibrium Statistical Mechanics. \textit{Phys. Rev. Lett.}, \textbf{74}, 2694-2697 (1995).

\bibitem{Evans1994} Evans, D.~J. and Searles, D.~J. Equilibrium microstates which generate second law violating steady states. \textit{Phys. Rev. E}, \textbf{50}, 1645-1648 (1994).

\bibitem{Evans1993} Evans, D.~J. and Cohen, E.~G.~D. and Morriss, G.~P. Probability of Second Law Violations in Shearing Steady States. \textit{Phys. Rev. Lett.}, \textbf{71}, 2401-2404 (1993).

\bibitem{Jarzynski2004} Jarzynski, C., \& W\'ojcik, D.~K. Classical and Quantum Fluctuation Theorems for Heat Exchange. \textit{Phys. Rev. Lett.}, \textbf{92}, 230602 (2004).\\
\end{thebibliography}
\end{document}